\documentclass{PoS}

\newcommand{\beq}{\begin{equation}}
\newcommand{\eeq}{\end{equation}}
\newcommand{\bea}{\begin{eqnarray}}
\newcommand{\eea}{\end{eqnarray}}
\newcommand{\bmp}{\begin{minipage}}
\newcommand{\emp}{\end{minipage}}
\newcommand{\D}{\displaystyle}

\newcommand{\tr}{{\rm tr}}

\newcommand{\V}{{\cal V}}

\input epsf

\PoS{PoS(LAT2005)311}

\title{RG decimation-based approach to confinement and computation on 
coarser lattices }

\ShortTitle{RG decimation-based approach to confinement}

\author{\speaker{E.T.~Tomboulis}\\
        Univ of California, Los Angeles\\
        E-mail: \email{tombouli@physics.ucla.edu}}

\abstract{A systematic procedure is presented for connecting short to 
long scales in LGT. Approximate decimations are used which can provide 
both upper and lower bounds on the  partition function. Its exact 
value is then obtained by interpolation between the bounds. By iterating 
the procedure representations of the partition function as well as other  
physical quantities can be obtained on progressively coarser lattices. 
For $SU(2)$ IR flow into the confining strong coupling regime results 
for any initial $\beta$.}

\FullConference{XXIIIrd International Symposium on Lattice Field Theory\\
		 25-30 July 2005\\
		 Trinity College, Dublin, Ireland}

\begin{document}

\section{Introduction}
As it is well-known, pure non-abelian  
gauge theory at $T=0$ and space-time dimension $d\leq 4$ possesses 
the extraordinary property of being in a single phase for all values of 
the bare gauge coupling, $ 0 < \beta <\infty$. 
The theory exhibits passage  
from the short distance perturbative (ordered) regime to a long distance 
confining (disordered) regime without encountering a phase transition.  
Such multi-scale problems are notoriously difficult to treat from 
first principles. Another well-known, and in fact somewhat similar  
example is the passage from  laminar to turbulent flow in hydrodynamics. 

In principle such problems can be treated by some 
systematic block-spinning or decimation procedure.   
Exact block-spinning schemes in gauge theories, however,  so 
far appear intractable, both analytically and numerically. 
There are, however, approximate but computable decimation procedures that 
can provide bounds on judicially chosen quantities. Such bounds 
can then be used to rigorously constrain the flow behavior of the exact 
theory \cite{T1} - \cite{T2}. 
The basic procedure followed in this program is as follows: 

(i) Devise decimation schemes such that at each 
decimation step one can obtain  both an upper {\it and} a 
lower bound for the partition function. Furthermore, the decimations should 
be explicitly computable to any accuracy. 

(ii) At each step interpolate between the upper and lower 
bound. Then the exact value of partition partition function 
(free energy) is obtained at some particular value of the 
interpolating parameter. 
This then gives a {\it representation} of the partition  function 
on the decimated lattice \cite{T1}.   

(iii) The same procedure is then applied to the partition function 
in the presence of a `twist' (external center flux).

(iv) Combining (ii) and (iii) allows consideration of appropriate 
long distance order parameters: the vortex free energy, Wilson loop, 
and Wilson lines.

We will see that the flow into the IR regime of the exact theory 
can be extracted {\it without} knowing the precise  right numerical 
values of the extrapolation parameters that give the exact 
representations at each decimation step. For $SU(2)$, the only 
case explicitly considered here, the result is  
flow into the strong coupling confinement regime for any initial $\beta$.
Furthermore, if these right numerical values of the extrapolation parameters
can be explicitly obtained, one can compute appropriate quantities directly on 
coarser lattices.  
 
\section{Decimation and interpolation}
Starting  with some reflection positive 
plaquette action at lattice spacing $a$, e.g. the Wilson action  
$A_p(U) ={\beta\over N}\;{\rm Re}\,\tr U_p$,  
consider the character expansion of the action exponential:
\beq
F(U, a) =e^{A_p(U)} 
   = \sum_j\;d_j\,F_j(\beta,a)\,\chi_j(U) \;,
\eeq
For SU(2), one has  
$j=0, {1\over 2}, 1, {3\over 2}, \ldots$, $d_j=(2j+1)$.
It is convenient to work in terms of normalized coefficients; 
\beq
F(U, a) =  F_0\,\Big[\, 1 + \sum_{j\not= 0} c_j(\beta)\,d_j\,\chi_j(U)\,
 \Big] 
 \equiv  F_0\;f(U,a)\;.  
\eeq 
The partition function on lattice $\Lambda$ is then 
\beq
Z_\Lambda(\beta) = \int dU_\Lambda\;\prod_p\,F_p(U,a)
                    =F_0^{|\Lambda|}\; \int dU_\Lambda\;\prod_p\,f_p(U,a)\;.
\eeq

We employ decimation schemes of the `potential moving' type. The 
lattice is split into hypercubes of side length $\lambda a$, and 
plaquettes are moved from the interior of each hypercube to 
its boundary, where plaquette interactions are appropriately adjusted to 
compensate for the move. 
The process can be summarized as an explicit decimation 
rule for each step in successive decimations
\[ a \to \lambda a \to \lambda^2 a \to \cdots \to \lambda^n a 
\qquad \Longleftrightarrow \qquad 
 \Lambda \to \Lambda^{(1)} \to \Lambda^{(2)} \to \cdots \to 
\Lambda^{(n)} \;,\]
which gives  explicit expressions for the computation of the Fourier 
coefficients at the $(m+1)$-th step given those of the $m$-th step: 
\[ F_0(m) \to F_0(m+1) = F_0(m, \zeta, r, \lambda, \{ c_i(m)\})\]
\[ c_j(m) \to c_j(m+1)= c_j(m, \zeta, r, \lambda, \{ c_i(m)\})  \] 

The rule for the decimation scheme we adopt, which need not be 
explicitly given here, involves parameters 
$\zeta$, $r$ which control the amount 
by which the interactions of the remaining plaquettes after a decimation 
step are renormalized to compensate for the ones that were removed. 

The resulting PF after $n$ decimation steps is then: 
\beq 
Z_\Lambda(\beta, n) 
                    =\prod_{m=0}^n F_0(m)^{|\Lambda|/\lambda^{md}}\; 
\int dU_{\Lambda^{(n)}}\;\prod_p\,f_p(U,n)  \;.
\eeq
It is important to note that, after each decimation step, the resulting action 
retains the original {\it one-plaquette form} but will, in general,  
contain all representations:
\beq
f_p(U,n) =  \Big[\, 1 + \sum_{j\not= 0} c_j(n)\,d_j\,\chi_j(U)\,
 \Big] \equiv  \exp A_p(n) \label{action} 
\eeq
with   
\beq
A_p(n)= \sum_j \; \tilde{\beta}_j(\beta, n)\,\chi_j(U) \;.
\eeq
Also, both positive and negative effective couplings 
$\tilde{\beta}_j$ will, in general,  occur.

The following two basic statements can now be proved. 
Going from the $n-1$ step to the $n$ decimation step one has:

(I) with $\zeta=\lambda^{d-2}$, $\quad r=1-\epsilon\;, \qquad 
0\leq \epsilon < 1$ :  
\beq
Z_\Lambda(\beta, n-1) \leq Z_\Lambda(\beta, n) \;;
\eeq

(II) with $\zeta=1$, $\quad r=1+\epsilon \;, \qquad 
0\leq \epsilon$ : 
\beq
Z_\Lambda(\beta, n) \leq Z_\Lambda(\beta, n-1) \; .
\eeq 
In fact, in both (I), (II) one has strict inequality at any finite $\beta$.  

Translation invariance, convexity of the free energy, and 
reflection positivity (positivity of Fourier coefficients) underlie 
(I), (II). 

We next interpolate between the upper and lower bounds. 
Introducing a parameter 
$\alpha$, ($0\leq \alpha$), define  
interpolating coefficients $\tilde{c}_j(n,\alpha)$ and 
$\tilde{F}_0(n,\alpha)$ 
such that 
\[ \tilde{c}_j(n,\alpha)= \left\{ \begin{array}{lll}
c_j(n, \zeta=\lambda^{d-2}, r=1-\epsilon) & : &\alpha=1 \\ 
c_j(n, \zeta=1, r=1+\epsilon)  & : & \alpha=0 
\end{array} \right. \] 
and similarly 
\[ \tilde{F}_0(n,\alpha)= \left\{ \begin{array}{lll}
F_0(n, \zeta=\lambda^{d-2}, r=1-\epsilon) & : &\alpha=1 \\ 
F_0(n, \zeta=1, r=1+\epsilon) & : & \alpha=0 
\end{array} \right. 
\;\]
for $0\leq \epsilon < 1$.   
A variety of convenient explicit examples of such interpolating 
$\tilde{c}_j(n,\alpha)$, $\tilde{F}_0(n,\alpha)$ may be given. 
There is, of course, nothing unique about any one such smooth 
interpolation. Consider more generally a family of smooth interpolations 
parametrized by a parameter $t$ in some interval  $(t_1, t_2)$: 
\[ \tilde{c}_j(n,\alpha, t) \;,\qquad  \tilde{F}_0(n,\alpha,t) \]
Define the corresponding interpolating partition function 
$\tilde{Z}_\Lambda(\beta, n, \alpha,t)$
by the replacement 
\[ F_0(n) \to \tilde{F}_0(n,\alpha,t) \;, \qquad c_j(n) \to 
\tilde{c}_j(n,\alpha,t)  \] 
Then, for each fixed $t$, we have 
\beq
\tilde{Z}_\Lambda(\beta, n,0, t) \leq Z_\Lambda(\beta, n-1) 
\leq \tilde{Z}_\Lambda(\beta, n, 1, t)  \;.
\eeq
But then, {\it by continuity, there exist a value} 
\[ 0 < \alpha=\alpha^{(n)}_\Lambda(t) < 1 \] 
{\it such that} 
\beq
\tilde{Z}_\Lambda(\beta, n, \alpha^{(n)}_\Lambda(t), t)= 
Z_\Lambda(\beta, n-1) \;. \label{fix}
\eeq
In other words, $\alpha^{(n)}_\Lambda(\beta, t)$ is the level surface 
(implicit function) solution to 
\beq
\tilde{Z}_\Lambda(\beta, n, \alpha, t)= 
Z_\Lambda(\beta, n-1) \;.
\eeq 
There is thus `parametrization invariance' in (\ref{fix}): shifts in $t$ 
are compensated by changes in $\alpha^{(n)}_\Lambda(\beta, t)$.  
This procedure of fixing $\alpha^{(n)}_\Lambda(\beta, t)$ 
is iterated in successive decimations ($n=1,2,\ldots$).    
So starting at the original spacing $a$, after the 1st decimation step 
the procedure gives {\it an exact representation} on the coarser 
lattice $\Lambda^{(1)}$ of the original 
partition function on $\Lambda$ in the form:
\bea    
Z_\Lambda(\beta) &= & 
F_0^{|\Lambda|}\; \int dU_\Lambda\;\prod_p\,f_p(U,a) \nonumber \\
  & = & \tilde{Z}_\Lambda(\beta, \alpha^{(1)}_\Lambda(t), t) \nonumber \\
& = & F_0^{|\Lambda|}\,
\tilde{F}_0(\alpha_\Lambda^{(1)}(t),t)^{|\Lambda|/\lambda^{d}}\;  
\int dU_{\Lambda^{(1)}}\;\prod_p\,f_p(U,\alpha_\Lambda^{(1)}(t),t) 
\label{A} \;. 
\eea
Since the resulting action in (\ref{A}) is again of the same 
type (\ref{action}),  iteration of the procedure 
results in a representation analogous to (\ref{A}) on successively  
coarser lattices $\Lambda^{(n)}$.

Note that, by construction, the coefficients $ \tilde{c}_j$,  
$\tilde{F}_0$ evaluated at $\alpha_\Lambda^{(n)}(t)$ occurring in 
(\ref{A}) 
are always sandwiched between the 
explicitly computable upper bound ($\alpha=1$)  and lower bound  
($\alpha=0$) coefficients.

\section{Twisted partition function -- Vortex free-energy}
 
The twisted partition function $Z_\Lambda^{(-)}(\beta)$ is defined  
as the partition function with `twisted' action on each 
plaquette of a coclosed set ${\cal V}$ of plaquettes winding 
around the periodic lattice in $(d-2)$ directions perpendicular to a given 
$2$-plane. Thus on the dual lattice ${\cal V}$ forms a closed $2$-dim 
surface in $d=4$, and a closed loop in $d=3$. 
The `twist' is a shift by a non-trivial  element $\tau \in Z(N)$ 
of the group center,  and 
amounts to a discontinuous (singular) gauge
transformation on the configurations in the partition function
with multivaluedness in $Z(N)$, i.e. the introduction of a
$\pi_1(SU(N)/Z(N)) = Z(N)$ vortex. $\tau=-1$ for $SU(2)$. 

The vortex free-energy order parameter $F_\Lambda^{(-)}$ is then defined as: 
\beq 
\exp{-F_\Lambda^{(-)}}  = {Z_\Lambda^{(-)}\over Z_\Lambda} \label{vfe}\;.
\eeq
The behavior of the Wilson loop and Wilson line correlations can be 
related to (\ref{vfe}) through known inequalities. 

There is a slight technical complication in applying the above procedure 
to the twisted partition function.   
For $Z_\Lambda^{(-)}$ reflection positivity holds only in planes 
perpendicular to the directions in which the set $\V$ carrying the twist is 
winding around the lattice. One way to overcome this is to 
simply replace $Z_\Lambda^{(-)}$ by the quantity 
$Z_\Lambda + Z_\Lambda^{(-)}$ for which reflection positivity, as it 
is easily verified, is restored in all planes. The above 
development then can be carried through in this case also. 

Thus, e.g., after one decimation, one has  the 
{\it exact representation} analogous to (\ref{A}): 
\beq
Z_\Lambda(\beta) + Z_\Lambda^{(-)}(\beta)  =  
\tilde{Z}_\Lambda\Big( \beta,  
\alpha_\Lambda^{+(1)}(t),t \Big) 
  + \tilde{Z}_\Lambda^{(-)}\Big( \beta,  
\alpha_\Lambda^{+(1)}(t),t \Big)  \;. \label{B} 
\eeq
One may then iterate the procedure in successive decimation steps. 

It should be noted that, as indicated by the notation, the 
$\alpha_\Lambda^{+(1)}(t)$ in (\ref{B}) are fixed independently of and hence 
a priori can be distinct  from the 
$\alpha_\Lambda^{(1)}(t)$ occurring in the representation (\ref{A}) 
for the partition function $Z_\Lambda(\beta)$. The same of course is true 
for each subsequent iteration.

Consider now using these exact representations for computing the 
vortex free-energy (\ref{vfe}). 
After one decimation, using (\ref{B}), one has: 
\bea
\left( 1 + {\D Z_\Lambda^{(-)}\over \D Z_\Lambda}\right) 
=  {Z_\Lambda + Z_\Lambda^{(-)} \over Z_\Lambda}
& = & { \tilde{Z}_\Lambda\Big( \beta, 
\alpha_\Lambda^{+(1)}(t), t\Big) 
+ \tilde{Z}_\Lambda^{(-)}\Big( \beta,  
\alpha_\Lambda^{+(1)}(t),t \Big)
\over Z_\Lambda }
\label{vfe1}\\
& = & { \tilde{Z}_\Lambda\Big(  
\alpha_\Lambda^{+(1)}(t), t\Big) \over Z_\Lambda }
\Bigg[1 + { \tilde{Z}_\Lambda^{(-)}\Big(   
\alpha_\Lambda^{+(1)}(t),t \Big) \over 
\tilde{Z}_\Lambda\Big(  \alpha_\Lambda^{+(1)}(t), t\Big)}  \Bigg] 
\label{vfe2} \;.
\eea 
By construction, the numerator  
in (\ref{vfe1}) is invariant under shifts in $t$; 
but the two factors in which it is split in 
(\ref{vfe2}) are {\it not} individually invariant. 
(They would be, by (\ref{A}), only if $\alpha_\Lambda^{+(1)}(t)$ were to 
exactly coincide with $\alpha_\Lambda^{(1)}(t)$ for all $t$.)  

One now finds that there exist a value $t=t^*$ at which  
$\tilde{Z}_\Lambda\Big(  \alpha_\Lambda^{+(1)}(t), t \Big)$ intersects  
the level curve $Z_\Lambda 
=\tilde{Z}_\Lambda\Big(\alpha_\Lambda^{(1)}(t), t\Big)$, i.e. 
\[ \tilde{Z}_\Lambda\Big(  
\alpha_\Lambda^{+(1)}(t^*), t^*\Big) = Z_\Lambda 
=\tilde{Z}_\Lambda\Big(\alpha_\Lambda^{(1)}(t), t\Big) \;. \] 
Thus 
\beq
1 + {\D Z_\Lambda^{(-)}\over \D Z_\Lambda} 
= 1 + { \tilde{Z}_\Lambda^{(-)}\Big(   
\alpha_\Lambda^{+(1)}(t^*),t^* \Big) \over
\tilde{Z}_\Lambda\Big(  
\alpha_\Lambda^{+(1)}(t^*), t^*\Big)}  \;.\label{vfe3}
\eeq
(\ref{vfe3}) may then be iterated in successive decimation steps. 
Now, by construction, the coefficients  
$\tilde{c}_j(n,\alpha_\Lambda^{+(n)}(t))$, 
$\tilde{F}_0(n,\alpha_\Lambda^{+(n)}(t),t)$ are sandwiched between the 
upper bound ($\alpha=1$)  and lower bound 
($\alpha=0$) coefficients which, for the 
potential-moving type of decimations employed here, are explicitly 
computable. In particular, the  upper bound ones are essentially those of 
a Migdal-Kadanoff decimation. For $SU(2)$ this implies flow 
of the effective action in (\ref{vfe3}) into the strong coupling regime, 
hence confining behavior for the order parameter,  
for any initial $\beta$.   
Decimations need only be carried out till the strong coupling 
regime is reached. From that point onward, computations 
can be performed within the convergent strong coupling cluster 
expansion.

\section{Conclusion -- Outlook} 
The iterated procedure of decimation followed by interpolation 
between computable upper and lower bounds allows one to 
constrain and hence follow 
the qualitative RG flow of the exact theory. This is possible 
without knowledge of  the precise  
numerical values of the extrapolation parameters (the 
$\alpha_\Lambda^{(n)}$'s) 
needed to give the exact representation of the physical quantity  
(partition functions, vortex free energy, etc) considered. 
For $SU(2)$, the case explicitly treated, this 
suffices to extract IR flow into the 
strong coupling confinement regime for any initial $\beta$.

The derivations above only ascertain, within 
upper and lower bounds, the existence of 
appropriate values (the $\alpha^{(n)}$'s) for the interpolating 
parameters reproducing a given physical quantity on successively coarser 
lattices. If these exact numerical values of the extrapolation parameters
can be obtained, one can compute this quantity  directly on 
coarser lattices.    
An advantage of the above development is that the general form of the 
effective action entering at each step is determined (eq. (\ref{action})), and 
only the numerical coefficients (effective 
couplings) are to be determined. Algorithms for doing this, involving 
an improved microcanonical demon technique, are  
under development in collaboration with A. Velytsky. 

\bigskip

\noindent{\bf \large Acknowledgments:} This work was partially supported by 
NSF-PHY-0309362.


\begin{thebibliography}{99}
\bibitem{T1} E.T. Tomboulis, {\it RG decimations and confinement}, 
Nucl. Phys. B (Proc. Suppl.) {\bf 141} (2005) 115 [hep-lat/0409019].
\bibitem{T2} E.T. Tomboulis, {\it From short to long scales in the QCD 
vacuum}, Nucl. Phys. B (Proc. Suppl.) {\bf 129} (2004) 724 [hep-lat/0309006].
\end{thebibliography}
\end{document}